\documentclass[11pt]{article}
\usepackage{graphicx}

\newsavebox{\hflrar}
\sbox{\hflrar}{\makebox[0pt][l]
{${\scriptstyle \leftharpoonup}$}{${\scriptstyle \rightharpoonup}$}}

\def \to {\rightarrow}

\def\bfsig{\mbox{\boldmath$\sigma$}}

\begin{document}
\pagestyle{plain}
\vskip 10mm
\begin{center}
{\bf\Large NRQCD Factorization for Twist-2 Light-Cone Wave-Functions of Charmonia} \\
\vskip 10mm
J.P. Ma$^{1,2}$ and Z.G. Si$^{2}$   \\
{\small {\it $^1$Institute of Theoretical Physics , Academia
Sinica, Beijing 100080, China \ \ \ }} \\
{\small {\it $^2$ Department of Physics, Shandong University, Jinan Shandong 250100, China}}
\end{center}
\vskip 0.4 cm

\begin{abstract}
We show that the twist-2 light-cone wave-functions of $\eta_c$ and $J/\psi$ can be factorized
with nonrelativistic QCD(NRQCD) at one-loop level, where the nonperturbative effects are represented
by NRQCD matrix elements.
The factorization is achieved by expanding the small velocity $v$, which the $c$- or $\bar c$- quark
moves inside a rest quarkonium with.
At leading order of $v$ the twist-2 light-cone wave-functions of $\eta_c$ and $J/\psi$ can be factorized
as the product of a perturbative function and a NRQCD matrix element. The perturbative function
is calculated at one-loop level and free from any soft divergence. Our results can be used
for the production of $J/\psi\eta_c$ through $e^+e^-$ -annihilation and of a charmonium
in $B$-decays, which are studied in experiment of two $B$-factories.

\end{abstract}
\par\vfil\eject

\par
Exclusive processes involving a charmonium can be studied with nonrelativistic QCD(NRQCD)\cite{NRQCD}.
One uses the fact that the $c$- or $\bar c$-quark moves with a small velocity inside a
rest charmonium and makes an expansion in $v$ to factorize the perturbative- and nonperturbative
effects. In this approach of NRQCD factorization
the production of a charmonium can be viewed as a two-step process,
where a free $c\bar c$-pair is first produced and then the pair is transformed into the
bound state of the charmonium. Because the charm quark mass $m_c$ is large, the production
of the $c\bar c$-pair can be calculated with perturbative QCD. The transformation is nonperturbative
and can be characterized with NRQCD matrix elements. Applications of the approach can be
found in the comprehensive reveiw\cite{YellowBook}.
\par
Although the approach of NRQCD factorization is consistent in theory, but in practices
there are problems when confronting experiments. First,
the expansion in $v$ converge slowly because $v$ for charmonia is not small enough
with the typical value $v^2 \sim 0.3$.
Secondly, since in the approach the charm quark mass
must be kept as nonzero, there will be in the perturbative expansion
large logarithms terms like $\ln(m_c/Q)$, when
the process of the production involves some large scales, generically denoted as $Q$, and $Q>>m_c$.
These large logarithms terms will spoil the applicability of perturbative QCD.
\par
Alternatively, one can use the collinear factorization for those exclusive processes
involving some large scales with $Q>>m_c$.  It should be noted that the collinear factorization
for exclusive processes of light hadrons has been suggested long time ago\cite{BL,CZrep}.
In this approach the nonperturbative effects are parameterized with
different light-cone wave-functions(LCWF's) of a charmonium and the amplitude
of a process can be written as a convolution with those LCWF's and a perturbative function.
In calculating the perturbative function charm quarks
should be taken as light quarks at leading twist, i.e., $m_c=0$. Therefore, large logarithms
terms $\ln(m_c/Q)$ will not appear in the perturbative function, but will appear
in LCWF's. Then those large logarithms terms can be easily re-summed by using
renormalization group evolutions of LCWF's. The situation here is similar
to inclusive production of a charmonium, where one uses the concept of fragmentation
functions to resum large logarithms terms.
\par
Experimentally exclusive production of a charmonium in $B$-decays have been studied
extensively at two $B$-factories.   An interesting study at $B$-factories reveals a puzzle
for theory. The cross-section of $e^+e^- \to J/\psi \eta_c$ has been measured\cite{B-exp}.
The measured cross-section is about an order of magnitude larger than theoretical predictions.
The theoretical predictions are based more or less on the approach of NRQCD factorization\cite{NRth}.
Adding one-loop correction in the approach\cite{NRth1L}, the discrepancy still remains
although the correction is significant. It has been suggested\cite{MaSi,BC, BLL} that
by using the collinear factorization it is possible to solve the puzzle. In this approach
the nonperturbative effects related to charmonia are parameterized with different
LCWF's of charmonia. The collinear factorization has been also used in other exclusive processes
involving quarkonia\cite{Others}.
\par
In this letter we point out that LCWF's of charmonia can be studied with NRQCD,
in which one can factorize perturbative- and nonperturbative effects. The perturbative
effects are related to the fact that charm quarks are heavy. The situation is similar
to parton fragmentation functions into a charmoium for inclusive production, where
a NRQCD factorization can be performed\cite{Frag}. We will show here that the NRQCD factorization
for LCWF's can be performed at one-loop level for twist-2 LCWF's of $J/\psi$ or $\eta_c$, i.e., of
a $S$-wave quarkonium, where we make an expansion in $v$ and only keep the leading order.
At the order the shape of LCWF's is completely determined by perturbative QCD.
It should be noted that LCWF's of charmonia have been studied in a potential model\cite{BKL}.
\par
We will use the  light-cone coordinate system, in which a
vector $a^\mu$ is expressed as $a^\mu = (a^+, a^-, \vec a_\perp) =
((a^0+a^3)/\sqrt{2}, (a^0-a^3)/\sqrt{2}, a^1, a^2)$ and $a_\perp^2
=(a^1)^2+(a^2)^2$. There is one LCWF of $\eta_c$ at leading twist,
which is defined as:
\begin{equation}
\phi_{\eta_c} (u, \mu) =\int \frac{dx^-}{2\pi} e^{+iuP^+ x^-}
\langle 0 \vert \bar c (0)L_n^\dagger (\infty, 0)
\gamma^+ \gamma_5  L_n(\infty, x^-n) c (x^- n) \vert \eta_c (P) \rangle,
\end{equation}
where $\eta_c$ carries the momentum $P^\mu =(P^+, P^-,0,0)$. $L_n$ is the gauge
link along the light-cone direction $n^\mu=(0,1,0,0)$ to make the definition gauge-invariant:
\begin{equation}
L_n (\infty, z) = P \exp \left ( -i g_s \int_0^{\infty} d\lambda
     n\cdot G (\lambda n + z) \right ) .
\end{equation}
The variable $u$ is the momentum fraction carried by the $c$-quark, i.e., the $c$-quark
has the $+$-component $uP^+$ of its momentum.
Similarly, the LCWF's of $J/\psi$ with different polarizations
$\lambda$ are defined as:
\begin{eqnarray}
\phi_{\|} (u, \mu) &=&\int \frac{dx^-}{2\pi} e^{+iuP^+ x^-}
\langle 0 \vert \bar c (0)L_n^\dagger (\infty, 0)
\gamma^+   L_n(\infty, x^-n) c (x^- n) \vert J/\psi (P, \lambda =0) \rangle,
\nonumber\\
\phi_{\perp} (u, \mu) &=& \int \frac{dx^-}{2\pi} e^{+iuP^+ x^-}
\langle 0 \vert \bar c (0)L_n^\dagger (\infty, 0)
 \left (-i\sigma^{+ \mu} \varepsilon_\mu^*(\lambda_\perp) \right )  L_n(\infty, x^-n) c (x^- n)
 \vert J/\psi (P, \lambda_\perp) \rangle,
\end{eqnarray}
where $\varepsilon_\mu$ is the polarization vector of $J/\psi$
and $n\cdot\varepsilon_\mu^*(\lambda =\pm 1)= l\cdot\varepsilon_\mu^*(\lambda =\pm 1)=0$.
($\varepsilon_\mu^*(\lambda =\pm 1) =\varepsilon_\mu^*(\lambda_\perp)$.
There are only two LCWF's of $J/\psi$ at leading twist.
These definitions can be easily obtained from those of light hadrons\cite{wfpion,wfrho},
the definitions here are only different in the normalization than those defined in \cite{wfpion,wfrho}.
The LCWF's  are Lorentz-invariant along the $z$-direction.
\par
In the framework of NRQCD based on the small velocity $v$ expansion, one expects that those functions
take a factorized form:
\begin{equation}
\phi_{\eta_c, \|, \perp} (u,\mu) = \hat \phi_{\eta_c, \|, \perp} (u,\mu)
       \frac{ \langle 0 \vert {\mathcal O}_H \vert H \rangle}{2\sqrt {m_c}} \cdot \left ( 1
        +{\mathcal O} (v^2) \right ),
\end{equation}
where $\hat \phi_n (n=\eta_c,\|,\perp)$ can be calculated in perturbative QCD and
they are free from any infrared- and Coulomb singularity, as we will show.
They are dimensionless. $H$ stands for $J/\psi$ or $\eta_c$.
The matrix element $\langle 0 \vert {\mathcal O}_H \vert H \rangle$ is defined with NRQCD operators,
whose definitions will be given later. They contain all nonperturbative effects.
The factor $2\sqrt{m_c}$ appears because the states
in the definitions of LCWF's are covariantly normalized, while the states in NRQCD matrix
elements are usually nonrelativistically normalized.
\par
To extract the perturbative function $\hat \phi_n$, we replace the quarkonium state
with a $c\bar c$-pair state. After the replacement, one can calculate LCWF's
in the left hand side of Eq.(4) and the matrix elements in the right hand side with
perturbative QCD and then extract $\hat \phi_n$. In such a calculation,
infrared- and Coulomb singularities can appear in the left hand side, but the same
will also appear in the matrix elements in the right hand side so that $\hat\phi_n$
is free from any soft divergence. This will be shown in our calculation below.
Because LCWF's are Lorentz invariant in the $z$-direction,
one can calculate LCWF's in the rest frame of the pair. In the rest frame the $c\bar c$-pair state
is specified with $\vert c(p_1) \bar c (p_2) \rangle$, the momenta are given by:
\begin{equation}
 {\bf p_1} = m_c{\bf v}, \ \ \ \ \ \ {\bf p_2} = -m_c{\bf v}, \ \ \ \ \  P= p_1+p_2
\end{equation}
For easy calculation we let the quarks move in the $z$-direction, i.e., ${\bf v} =(0,0,v)$
and $v>0$. This will not bring any problem for $S$-wave quarkonia at leading order of $v$.
For convenience we introduce the variable $\omega$
the following:
\begin{equation}
  \omega =\frac{p_1^+}{P^+}, \ \ \
  \ \ \ \  \omega = \frac{1}{2} + \frac{1}{2} v + {\mathcal O}(v^3),
\end{equation}
and denote
\begin{equation}
 \Gamma_n = \Gamma_{\eta_c, \|, \perp}
 = \left \{ \gamma^+\gamma_5,\  \gamma^+,\  -i\sigma^{+\mu}\varepsilon_\mu^* \right \}.
\end{equation}
We will use Feynman gauge in this work.
\par
We will always use the notation for a quantity $A$ as $A =A^{(0)} + A^{(1)} +\cdots$,
where $A^{(0)}$ denotes the tree-level contribution and $A^{(n)}$ denotes the $n$-loop
contribution. At tree-level we have:
\begin{eqnarray}
\phi^{(0)}_n (u) &=& \delta ( u-\omega)\cdot \frac{1}{P^+} \bar v(p_2) \Gamma_n u(p_1).
\end{eqnarray}
The product of the spinors in the above can be expressed with two-component NRQCD spinors
$\xi$ and $\eta$ in the limit $v\to 0$:
\begin{eqnarray}
\bar v(p_2) \gamma^+ \gamma_5 u(p_1) & = &  \sqrt{2} m_c \eta^\dagger \xi +{\mathcal O} (v^2)
   = \sqrt{2} m_c \langle 0 \vert \chi^\dagger \psi \vert c(p_1) \bar c(p_2) \rangle^{(0)} + {\mathcal O} (v^2),
\nonumber\\
\bar v(p_2) \gamma^+  u(p_1) &=& \sqrt{2} m_c \eta^\dagger  \sigma^3 \xi +{\mathcal O} (v^2)
= \sqrt{2} m_c \langle 0 \vert \chi^\dagger \sigma^3 \psi \vert c(p_1) \bar c(p_2) \rangle^{(0)} + {\mathcal O} (v^2),,
\nonumber\\
\bar v(p_2) \left ( -i \sigma^{+\mu} \epsilon^*_{\perp \mu}  \right ) u(p_1) &=&
\sqrt{2} m_c \eta^\dagger  \bfsig\cdot {\boldmath\epsilon^*_\perp }  \xi +{\mathcal O} (v^2)
= \sqrt{2} m_c \langle 0 \vert \chi^\dagger \bfsig\cdot {\boldmath\epsilon^*_\perp }
 \psi \vert c(p_1) \bar c(p_2) \rangle^{(0)} + {\mathcal O} (v^2),
\end{eqnarray}
where $\psi$ and $\chi$ are NRQCD field operators. $\psi$ annihilates a $c$-quark and
$\chi$ creates a $\bar c$-quark. With the above expression we can define two NRQCD matrix elements
relevant to our work:
\begin{equation}
\langle 0 \vert O_{\eta_c} \vert \eta_c \rangle = \langle 0 \vert \chi^\dagger \psi \vert \eta_c \rangle,
\ \ \ \ \
\langle 0 \vert O_{J/\psi}\vert J/\psi \rangle  =
\langle 0 \vert\chi^\dagger \frac{1}{3}\sum_{\lambda}\bfsig\cdot {\boldmath\epsilon}^* (\lambda)  \vert J/\psi (\lambda) \rangle.
\end{equation}
In the above the quarkonium is in the rest frame. With the above results we can specify the operators
in Eq.(4) with $\hat \phi_n$ determined at tree-level:
\begin{eqnarray}
\phi_{\eta_c} (u,\mu) &=& \hat \phi_{\eta_c} (u,\mu)
       \frac{ \langle 0 \vert {\mathcal O}_{\eta_c} \vert \eta_c \rangle}{2\sqrt {m_c}} \cdot \left ( 1
        +{\mathcal O} (v^2) \right ),
\nonumber\\
\phi_{\|,\perp} (u,\mu) &=& \hat \phi_{\|,\perp} (u,\mu)
       \frac{ \langle 0 \vert {\mathcal O}_{J/\psi} \vert J/\psi \rangle}{2\sqrt {m_c}} \cdot \left ( 1
        +{\mathcal O} (v^2) \right ),
\nonumber\\
\hat \phi_{\eta_c}^{(0)} &=& \hat\phi_{\|}^{(0)}(u,\mu)=\hat\phi_{\perp}^{(0)}(u,\mu) = \delta (u -1/2).
\end{eqnarray}
\par

\begin{figure}[hbt]
\begin{center}
\includegraphics[width=11cm]{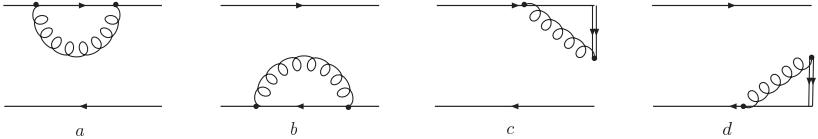}
\end{center}
\caption{The virtual part of the correction. The double lines
are for the gauge link.}
\label{dg1}
\end{figure}
\par
An important purpose of our work is to show that the perturbative functions can be
calculated with perturbative QCD at one-loop level and they are free from any soft
divergence. For this purpose we need to calculate $\phi_n$ and the NRQCD matrix elements
at one-loop to extract $\hat\phi_n$, where we take the partonic state $\vert c(p_1)\bar c(p_2)\rangle$.
As usual, the one-loop correction to $\phi_n$  can be divided into
a virtual- and a real part. The virtual part is represented
by diagrams given in Fig.1. The contributions from Fig.1a and Fig.1b
are from the external legs. They read:
\begin{equation}
\phi_n  {\vert}_{1a} (u)+\phi_n { \vert}_{1b} (u) = Z_W \phi^{(0)}_n  (u),
\end{equation}
with
\begin{equation}
Z_W =\frac{2\alpha_s}{3\pi} \left ( \frac{2}{\epsilon_I} +\gamma -\ln4\pi +\ln\frac{m_c^2}{\mu_I^2}
       + \frac{1}{2} \ln\frac{m_c^2}{\mu^2} -2 \right ),
\end{equation}
with $\epsilon_I = d-4$ for the infrared singularity and the scale $\mu_I$ for the singularity.
The U.V. poles in $\epsilon =4-d$ are subtracted.
The contribution from Fig.1c and Fig.1d reads:
\begin{eqnarray}
\phi_n{ \vert}_{1c}(u) &=&
  -\frac{2\alpha_s}{3\pi} \phi_n^{(0)} ( u)
 \Gamma(\frac{\epsilon}{2})
\left (\frac{4\pi \mu^2}{m_c^2}\right )^{\epsilon/2}
\left [ \frac{1}{\epsilon_I} -\frac{1}{1-\epsilon} \right ],
\nonumber\\
\phi_n{ \vert}_{1d}(u)
&=& -\frac{2\alpha_s}{3\pi} \phi_n^{(0)} ( u)
 \Gamma(\frac{\epsilon}{2})
\left (\frac{4\pi \mu^2}{m_c^2}\right )^{\epsilon/2}
\left [ \frac{1}{\epsilon_I} -\frac{1}{1-\epsilon} \right ],
\end{eqnarray}
Where $\Gamma(\epsilon/2)$ is for U.V. divergence and needs to be subtracted.
In the above there are also I.R. singularities. They will be canceled
by those in the real correction.
\par

\begin{figure}[hbt]
\begin{center}
\includegraphics[width=11cm]{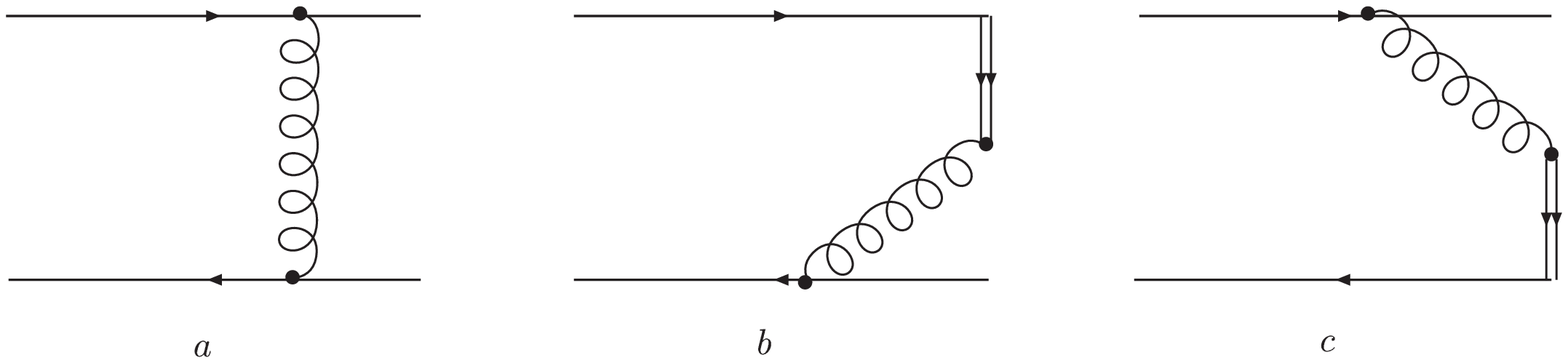}
\end{center}
\caption{The real part of the correction. The double lines
are for the gauge link.}
\label{dg2}
\end{figure}
\par
The real correction comes from diagrams
given in Fig.2. The contribution from Fig.2b and Fig.2c reads
\begin{eqnarray}
\phi_n{ \vert}_{2b}(u)
  &=& -\frac{2\alpha_s}{3\pi P^+}\bar v(p_2) \Gamma_n u(p_1)
 \Gamma(\frac{\epsilon}{2})
\left (\frac{4\pi \mu^2}{4 m_c^2}\right )^{\epsilon/2} \left [ 2 (u-\frac{1}{2})^{-\epsilon}
    -(u-\frac{1}{2})^{-1+\epsilon_I} \right ]\theta(u-1/2) +{\mathcal O}(v^2),
\nonumber\\
\phi_n{ \vert}_{2c}(u) &=&
-\frac{2\alpha_s}{3\pi P^+}\bar v(p_2) \Gamma_n u(p_1)
 \Gamma(\frac{\epsilon}{2})
\left (\frac{4\pi \mu^2}{4m_c^2}\right )^{\epsilon/2} \left [ 2(\frac{1}{2}-u)^{-\epsilon}
    -(\frac{1}{2} -u)^{-1+\epsilon_I} \right ]\theta(1/2-u)+{\mathcal O}(v^2).
\nonumber\\
\end{eqnarray}
In the above contributions there are poles at $ u=1/2$ when $\epsilon_I=0$.
These poles are infrared one and should be regularized as the poles
in $\epsilon_I$.
Using the expansion for $(\frac{1}{2} -u)^{-1+\epsilon_I}$ and
$(u-\frac{1}{2})^{-1+\epsilon_I}$ as distributions:
\begin{eqnarray}
(u-\frac{1}{2})^{-1+\epsilon_I}\theta(u-1/2) &=&
     \frac{1}{\epsilon_I} \left (\frac{1}{2}\right )^{\epsilon_I} \delta(u-1/2)
        + \theta(u-1/2) \left ( \frac {(u-1/2)^{\epsilon_I}}{u-1/2}\right )_+,
\nonumber\\
(\frac{1}{2}-u)^{-1+\epsilon_I}\theta(1/2-u) &=&
     \frac{1}{\epsilon_I} \left (\frac{1}{2}\right )^{\epsilon_I} \delta(u-1/2)
        + \theta(1/2-u) \left ( \frac {(1/2 -u)^{\epsilon_I}}{1/2-u}\right )_+,
\end{eqnarray}
with the $+$-distribution defined as:
\begin{eqnarray}
\int_{1/2}^1 d u f(u)\left ( \frac {(u-1/2)^{\epsilon_I}}{u-1/2}\right )_+
                  &=& \int_{1/2}^1 d u \left ( f(u) - f(1/2)\right )
                  \left ( \frac {(u-1/2)^{\epsilon_I}}{u-1/2}\right ),
\nonumber\\
\int^{1/2}_0 d u f(u)\left ( \frac {(1/2-u)^{\epsilon_I}}{1/2-u}\right )_+
                  &=& \int^{1/2}_0 d u \left ( f(u) - f(1/2)\right )
                  \left ( \frac {(1/2-u)^{\epsilon_I}}{1/2-u}\right ),
\end{eqnarray}
where $f(u)$ is a test function. With these results
one can show that the the infrared pole in $\epsilon_I$ is canceled
in the sum $\phi_n{ \vert}_{1c}+\phi_n{\vert}_{2c}$.
The same also happens for the sum $\phi_n{\vert}_{1d}+\phi_n{\vert}_{2b}$.
After subtracting the U.V. poles we have:
\begin{eqnarray}
 \phi_n{ \vert}_{1d}(u) +\phi_n{ \vert}_{2b}(u) &=& -\frac{2\alpha_s}{3\pi P^+} \bar v(p_2) \Gamma_n u(p_1)
\left \{ \ln\frac{\mu^2}{m_c^2} \left [ -\delta (u-1/2) -\theta(2u-1)\left (\left (\frac{2}{2u-1}\right )_+
 -2  \right ) \right ] \right.
\nonumber\\
  && \left.  -2 \delta(u-1/2) - 4\theta(2u-1)\left [ \ln (2u-1) - \left ( \frac{\ln(2u-1)}{2u-1}\right )_+
   \right ] \right \},
\nonumber\\
\phi_n{ \vert}_{1c}(u) +\phi_n{ \vert}_{2c}(u) &=&
\phi_n{ \vert}_{1d}(1-u) +\phi_n{\vert}_{2b}(1-u)
\end{eqnarray}
\par
The most difficult part is the contribution from Fig.2a.
It contains not only U.V.- and I.R. poles, but also the Coulomb singularity when $v$ goes to zero,
represented as $1/v$. After a tedious calculation we have:
\begin{eqnarray}
\phi_n{ \vert}_{2a}(u) &=& \frac{2 \alpha_s}{3\pi P^+} \bar v(p_2) \Gamma_n u(p_1)
             \left [ \delta (u-\omega) \left (
          -\frac{2}{\epsilon_I} -\gamma +\ln\frac{4\pi \mu_I ^2}{m_c^2} + \frac{\pi^2}{2v}
           -2 \right ) -2 \hat I(u) \right ] +\phi_n{ \vert}_{2a,R}(u),
\nonumber\\
\end{eqnarray}
where $\phi_n{ \vert}_{2a,R}(u)$ contains a U.V. pole and its form depends on the structure
of $\Gamma_n$. The contributions with the I.R. and Coulomb singularity are universal, they
do not depend on $\Gamma_n$. The distribution $\hat I(u)$ is defined as:
\begin{eqnarray}
\hat I (u) &=& \frac{\theta(1-2u)}{(1-2u)_+} -\frac{\theta(1-2u)}{(1-2u)^2_+}
 +\frac{\theta(2u-1)}{(2u-1)_+} -\frac{\theta(2u-1)}{(2u-1)^2_+},
\nonumber\\
\int_0^1 d u \frac{\theta(1-2u)}{(1-2u)^2_+} f(u) & = & \int_0^{1/2} d u\frac{f(u)-f(1/2)+(1/2 -u) f'(1/2)}{(1-2u)^2},
\nonumber\\
\int_0^1 d u \frac{\theta(2u-1)}{(2u-1)^2_+} f(u) & = & \int_{1/2}^1 du \frac{f(u)-f(1/2)-(u-1/2)f'(1/2)}{(2u-1)^2},
\end{eqnarray}
where $f(u)$ is a test function and $f'(u) = df(u)/du$.
The remainder part is:
\begin{eqnarray}
\phi_{\eta_c}{ \vert}_{2a,R}(u) &=& \phi_{\| }{ \vert}_{2a,R}(u)
=\frac{4\alpha_s}{3\pi P^+}\bar v(p_2) \gamma^+ u(p_1)
  \left \{  \theta(1-2u) u \left [ \ln\frac{\mu^2}{m_c^2}-3  -2 \ln (1-2u) \right ]
\right.
\nonumber\\
 && \left.  + \theta(2u-1) (1-u) \left [ \ln\frac{\mu^2}{m_c^2}-3 -2 \ln (2u-1) \right ] \right \} ,
\nonumber\\
\phi_{\perp} {\vert}_{2a,R}(u) &=&  0.
\end{eqnarray}

\par

\begin{figure}[hbt]
\begin{center}
\includegraphics[width=8cm]{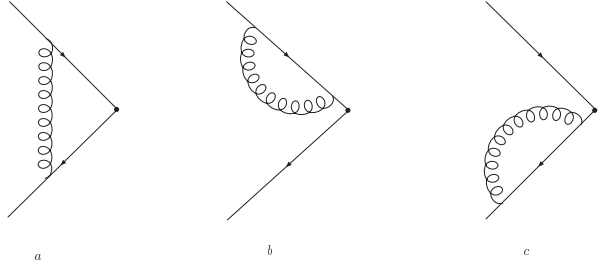}
\end{center}
\caption{One-loop correction for NRQCD matrix elements.} \label{dg2}
\end{figure}
\par
To verify the factorization in Eq.(11) one needs to calculate one-loop correction to
the NRQCD matrix elements. The one-loop corrections come from the diagrams given in
Fig.3. The calculation is straightforward. We have:
\begin{eqnarray}
\langle 0 \vert {\mathcal O}_H  \vert c({\bf v}) \bar c(-{\bf v})\rangle ^{(1)}
   &=& \frac{2\alpha_s}{3\pi}  \left [ - \left(\frac{2}{\epsilon_I}+\gamma-\ln \frac{4\pi \mu_I^2}{\mu^2} \right )
     +\frac{\pi^2}{2v} \right ]
      \langle 0 \vert  {\mathcal O}_H\vert c({\bf v}) \bar c(-{\bf v})\rangle^{(0)}
\nonumber\\
     &&   +\frac{2\alpha_s}{3\pi}\left (\frac{2}{\epsilon_I} +\gamma-\ln \frac{4\pi \mu_I^2}{\mu^2}\right )
  \langle 0 \vert {\mathcal O}_H \vert c({\bf v}) \bar c(-{\bf v})\rangle^{(0)} +{\mathcal O}(v^2),
\end{eqnarray}
The first line is for the contribution Fig.3a, the second line
is for the sum from Fig.3b and Fig.3c.
With above results we can now perform NRQCD factorization
and see how those I.R. singularities and Coulomb singularity are canceled.
\par
Comparing the one-loop results of NRQCD matrix elements with the one-loop
corrections to $\phi_n$, we can see that the Coulomb singularity and I.R. singularity from Fig.2a
in $\phi_n$ are canceled by the same singularities from Fig.3a in NRQCD matrix elements.
The I.R. singularities from Fig.1a and Fig.1b in $\phi_n$ are canceled by those
from Fig.3b and Fig.3c in NRQCD matrix elements. It should be noted
that the cancelations happen in the diagram-by-diagram way. This
leads to expect that the factorization can hold beyond one-loop level.
After these cancelations, the extracted
perturbative functions will be free from any soft divergence.
We have:
\begin{eqnarray}
\hat\phi^{(1)}_{\eta_c} (u,\mu) &=& \hat \phi^{(1)}_{\|} (u,\mu) = \frac{2\alpha_s(\mu)}{3\pi}
\left \{ \delta (u-1/2) \left ( 1 +\frac{3}{4} \ln\frac{\mu^2}{m_c^2} \right )
\right.
\nonumber\\
   && \left. + 2 \theta(1-2u)\left [ \ln\frac{\mu^2}{m_c^2}  \left ( \frac{1}{(1-2u)_+ } -1 +u \right )
   +   2(1-u)\ln (1-2u)
\right. \right.
\nonumber\\
  && \left.\left.  - 2\left ( \frac{\ln(1-2u)}{1-2u} \right )_+
     -\frac{1}{(1-2u)_+} +\frac{1}{(1-2u)^2_+} -3u\right ] + (u\to 1-u)  \right \},
\nonumber\\
\hat\phi^{(1)}_{\perp} (u,\mu) &=& \frac{2\alpha_s(\mu)}{3\pi}
\left \{ \delta (u-1/2) \left ( 1 +\frac{3}{4} \ln\frac{\mu^2}{m_c^2} \right )
\right.
\nonumber\\
   && \left. + 2 \theta(1-2u)\left [ \ln\frac{\mu^2}{m_c^2}  \left ( \frac{1}{(1-2u)_+ } -1 \right )
   +   2\ln (1-2u)
\right. \right.
\nonumber\\
  && \left.\left.  - 2\left( \frac{\ln(1-2u)}{1-2u } \right )_+
     -\frac{1}{(1-2u)_+} +\frac{1}{(1-2u)^2_+}\right ] + (u\to 1-u)  \right \}.
\end{eqnarray}
These are our main results.
\par
From our results we can also derive the dependence on the renormailization scale $\mu$.
For $\Gamma_n = \gamma^+$ and $\gamma^+ \gamma_5$:
\begin{eqnarray}
\frac {\partial \phi_{\eta_c,\|} ( u, \mu)}{\partial \ln \mu^2}
  = \frac{2\alpha_s}{2\pi} \left [ \frac{3}{4} \delta (1/2 -u) +2 \theta(1-2u) \left (
   \frac{1}{(1-2u)_+} + u-1 \right )
 +(u\to 1-u) \right ].
\end{eqnarray}
These LCWF's satisfy  the Efremov-Radyushkin-Brodsky-Lepage
evolution equation\cite{ERBL}. The
equation reads:
\begin{equation}
\frac{\partial \phi_{\eta_c, \|}(u,\mu)}{\partial \ln \mu^2}
= \int _0^1 dv V(u,v) \phi_{\eta_c, \|}  (v,\mu)
\end{equation}
with
\begin{equation}
V(u,v) =  \frac{2\alpha_s}{3\pi} \left [ \frac{ 1-u}{1-v}\left (1 +\frac{1}{u-v}\right )
\theta (u-v) +\frac{u}{v} \left ( 1+\frac{1}{v-u} \right ) \theta (v-u) \right ]_+
  +{\mathcal O}(\alpha_s^2).
\end{equation}
The $+$-prescription is
\begin{equation}
 [ V(u,v) ]_+ = V(u,v) - \delta(u-v) \int_0^1  dt V(t,v).
\end{equation}
Our results agree with the RG equation. Form our result it is easy
to derive the renormailzation group equation for $\phi_\perp$. We have
\begin{eqnarray}
\frac{\partial \phi_{\perp} (u,\mu)}{\partial \ln \mu^2}
& =&  \int _0^1 dv V_\perp (u,v) \phi_{\perp} (v,\mu),
\nonumber\\
V_\perp (u,v) &=&  V(u,v) - \frac{2\alpha_s}{3\pi} \left [
 \frac{u}{v}\theta(v-u) + \frac{1-u}{1-v} \theta (u-v) \right ]
\nonumber\\
  &=& \frac{2\alpha_s}{3\pi} \left [ \frac{3}{2} \delta (u-v)
     +\theta (v-u) \left ( \frac{1}{(v-u)_+} -1 \right ) + \theta (u-v) \left ( \frac{1}{(u-v)_+} -1 \right )\right ],
\end{eqnarray}
where the $+$-prescription acts on the variable $u$.
\par
To minimize possibly large perturbative corrections our results can be used
as initial conditions at $\mu=m_c$ to obtain LCWF's through renormalization
group equations at a scale $\mu$ relevant in a process. Through this step
the mentioned large logarithms terms $\ln (Q/m_c)$, mentioned at the beginning,
are resummed. In this work, the shape of LCWF's is completely determined
by perturbative QCD. Since $v^2$ is not small enough for charmonia,
nonperturbative effects can be at higher orders of $v^2$ can be important.
However, it is possible to resum these effects. E.g., one can consider
exchanges of multiple Coulomb glouns in Fig.2a. The resummation of these
exchanges leads to solve a Schr\"odinger equation. The resummation
results in that the sharp distribution $\delta(u-1/2)$ becomes smeared.
This can be studied with potential models as in \cite{BKL}.
\par
To summarize: We have shown that in the framework of NRQCD
the twist-2 LCWF's of a $S$-wave quarkonium
can be factorized as a product of perturbative functions and a NRQCD matrix element
in the nonrelativistic limit.
The nonperturbative effect is contained in the matrix element and the perturbative functions
can be calculated with perturbative QCD. They are determined at one-loop level
and free from I.R.- and Coulomb singularity. The LCWF's have a wide application
in exclusive production of quarkonia. Our results can be used as initial conditions
at $\mu =m_c$ to obtain the LCWF's at other scale $\mu$ relevant in a process.
With the study performed for twist-2 LCWF's we expect that such a factorization
also holds for twist-3 LCWF's. A detailed study of twist-3 LCWF's and phenomenological
applications of these results are under way.
\vskip20pt
\noindent
{\bf Acknowledgements}
\par
This work is supported by National Nature
Science Foundation of P. R. China.
\par\vfil\eject


\end{document}